\documentclass[12pt]{article}
\usepackage{amssymb}
\usepackage[dvips]{epsfig}

\newcommand{\be}{\begin{equation}}
\newcommand{\ee}{\end{equation}}
\def\bea{\begin{eqnarray}}
\def\eea{\end{eqnarray}}

\newcommand{\bn}{\begin{eqnarray}}
\newcommand{\en}{\end{eqnarray}}

\newcommand{\p}{\partial}

\newcommand{\nn}{\nonumber}
\newcommand{\tih}{\tilde{h}}

\newcommand{\tmnd}{T_{\left[\mu\nu\right]\rho}}
\newcommand{\tmnu}{T^{\left[\mu\nu\right]\rho}}
\newcommand{\no}{\noindent}

\newcommand{\s}{\,\,\,\,}
\def\bea{\begin{eqnarray}}
\def\eea{\end{eqnarray}}

\newcommand{\beq}{\begin{eqnarray}}
\newcommand{\eeq}{\end{eqnarray}}
\begin{document}

\title{\textbf{Nonuniqueness of the Fierz-Pauli mass term for a nonsymmetic tensor}}
\author{D. Dalmazi\footnote{dalmazi@feg.unesp.br} \\
\textit{{UNESP - Campus de Guaratinguet\'a - DFQ} }\\
\textit{{Avenida Dr. Ariberto Pereira da Cunha, 333} }\\
\textit{{CEP 12516-410 - Guaratinguet\'a - SP - Brazil.} }\\}
\date{\today}
\maketitle

\begin{abstract}

Starting with a description of massive spin-2 particles in $D=4$
in terms of a mixed symmetry tensor $\tmnd $ without totally
antisymmetric part ($T_{\left[\left[\mu\nu\right]\rho\right]}=0$)
we obtain a dual model in terms of a nonsymmetric tensor
$e_{\mu\nu}$. The model is of second-order in derivatives and its
mass term $\left( e_{\mu\nu}e^{\nu\mu} + c \, e^2\right)$ contains
an arbitrary real parameter $c$. Remarkably, it is free of ghosts
for any real value of $c$ and describes a massive spin-2 particle
as expected from duality. The antisymmetric part
$e_{\left[\mu\nu\right]}$ plays the role of auxiliary fields,
vanishing on shell. In the massless case the model describes a
massless spin-2 particle without ghosts.

\end{abstract}

\newpage

\section{Introduction}

Motivated  mainly by applications in the large scale gravitational
physics there have been intense work on infrared modifications of
gravity in the literature \cite{dgp}-\cite{hr12}, for review works
see \cite{rt08,h11}. Some ingenious solutions to two basic
problems of massive gravity, i.e., the appearance of ghosts
\cite{db} and the vDVZ mass discontinuity \cite{vdv,zak} have been
suggested based on \cite{vain}.

It is fair to say that the above problems are born in the free
massive Fierz-Pauli (FP) theory \cite{fp}. In particular, the
absence of ghosts in the free theory requires a fine tune $(c=-1)$
of the mass term $\left(h_{\mu\nu}^2 + c\, h^2\right)$ which
amounts to set the mass of the ghost to infinity. It is therefore
desirable to look for alternative descriptions of massive spin-2
particles. This is the subject of this work.

In terms of a symmetric tensor $h_{\mu\nu}=h_{\nu\mu}$, which is the minimal tensor structure required for a
spin-2 particle, the massive FP theory is unique\footnote{Up to trivial field redefinitions $h_{\mu\nu} \to
\tilde{h}_{\mu\nu}  + \frac a2 \tilde{h}\, \eta_{\mu\nu} $ with $a$ an arbitrary real number except $a=-1/2$
which is not invertible.} as a second-order theory. This point has been addressed in \cite{pvn73} and more
recently in \cite{abgv}. In \cite{abgv} starting with the massless case one notices that there is a whole
continuous family of theories which contains a massless spin-2 particle and is free of ghosts. They have been
named TDIFF Lagrangians. In general they also contain an extra scalar particle. There are only two points in the
parameters space where we get rid of the scalar field. One case is the popular massless FP theory (linearized
Einstein-Hilbert (LEH)) which might be called also DIFF theory and the other possibility is the WTDIFF
model\footnote{In the flat space TDIFF stands for transverse ($\p_{\mu}\xi^{\mu}=0$) linearized
reparametrizations $\delta h_{\mu\nu} = \p_{\mu}\xi_{\nu} + \p_{\nu}\xi_{\mu}$ while the WTDIFF model is
invariant also under linearized Weyl transformations $\delta_W h_{\mu\nu} = \phi \, \eta_{\mu\nu}$. The term
DIFF stands for unconstrained linearized reparametrizations. This explains one less unity in the number of
degrees of freedom in the WTDIFF and DIFF theories as compared to TDIFF.}. It turns out that only LEH can be
consistently (ghost-free) deformed in order to accommodate a massive spin-2 particle and that leads to the
uniqueness of the massive FP theory in terms of a symmetric tensor.

If we want to generalize the massive FP theory we may, for
instance, increase the number of symmetric tensors, allow for
nonsymmetric tensors or increase the rank of the tensor. In the
next section we start with the third possibility and end up with
the second one via a master action approach \cite{dj}. Remarkably,
we derive  a consistent model for a massive spin-2 particle in
terms of a nonsymmetric tensor which differs from the FP theory in
both the kinetic and massive terms. There is an arbitrariness in
the mass term which is not necessarily the usual Fierz-Pauli one.
The model is proved to be ghost-free via analysis of the analytic
structure of the propagator. In section 3 we study the massless
case and show that the gauge symmetries correspond to WDIFF (Weyl
and linearized reparametrizations) plus a reducible vector
symmetry in the antisymmetric sector, see (\ref{gt2}). In section
4 we draw our conclusions. In both sections 2 and 3 we analyze the
particle content via equations of motion and analytic structure of
the propagator.

\section{The massive case}

It is possible to formulate the Einstein-Hilbert gravity in a
first-order frame-like formalism in terms of the spin connection
$\omega_{\mu ab}=-\omega_{\mu ba} $ and the Vierbein $e_{\mu}^a$
treated as independent variables. An analogous formulation for a
massless spin-2 particle exists in the flat space. If we keep
using the curved space notation with only Greek indices and add a
Fierz-Pauli mass term for $e_{\mu\nu}$, the first-order Lagrangian
density in the flat space can be written symbolically (dropping
some indices) as ${\cal L}\left[ e, \omega \right] = -\omega^2 +
\omega \p e - m^2 \left(e_{\mu\nu}e^{\nu\mu} - e^2\right)$. If we
Gaussian integrate over the rank-3 tensor
$\omega_{\mu\left[\alpha\beta\right]}$ we derive the massive
Fierz-Pauli theory which describes a massive spin-2 particle in
terms of the nonsymmetric tensor $e_{\mu\nu}$. The antisymmetric
part $e_{\left[\mu\nu\right]}$ appears only in the mass term and
as such decouples trivially without any important contribution.
Instead, if we Gaussian integrate over $e_{\mu\nu}$ we end up with
a dual theory for the rank-3 tensor. It has been shown in
\cite{gkmu}, which makes use of \cite{zino}, that this higher-rank
description of a massive spin-2 particle in $D=4$ is the theory
suggested by Curtright \cite{curt,cf} in terms of a mixed symmetry
tensor $\tmnd $ which is a kind of dual spin connection. The
theory of \cite{curt,cf}, after a scaling by a mass factor, can be
conveniently written \cite{bfrt} as\footnote{Throughout this work
we use $\eta_{\mu\nu}=diag(-,+,+,+)$.}

\be {\cal L}_{C}\left[T\right] = E^{\mu\nu}_{\quad\beta} \tmnd
E^{\lambda\sigma\rho}T_{\left[\lambda\sigma\right]}^{\quad\beta} - 2\, m^2 \left\lbrack \tmnu \tmnd - 2\,
T^{\mu}T_{\mu} \right\rbrack \label{lc} \ee

\no where

\bea T_{\left[\mu\nu\right]\rho} &=& - T_{\left[\nu\mu\right]\rho}
\quad ; \quad T_{\mu} = \eta^{\nu\rho} \tmnd \label{t1} \\
T_{\left[\left[\mu\nu\right]\rho\right]} &=& \frac 13 \left(\tmnd
+ T_{\left[\nu\rho\right]\mu} + T_{\left[\rho\mu\right]\nu}\right)
= 0 \label{t2} \\
E^{\mu\nu\alpha} &=& \epsilon^{\mu\nu\alpha\rho}\p_{\rho}
\label{eo} \eea

\no Introducing\footnote{Although we use the same notation, the
new field should not be confused with $e_{\mu\nu}$ appearing in
${\cal L}\left[ e, \omega \right]$.} a nonsymmetric tensor
$e_{\mu\nu}$ we define the first-order Lagrangian density

\be {\cal L}\left[e,T\right] = - e_{\beta\rho} e^{\rho\beta} - c\,
e^2 + 2\, V_{\beta\rho}(T) e^{\rho\beta} - 2\, m^2 \left\lbrack
\tmnu \tmnd - 2\, T^{\mu}T_{\mu} \right\rbrack \label{let} \ee

\no where $c$ is an arbitrary real constant, $e=e_{\mu}^{\,\,\mu}$
and

\be V_{\beta\rho}(T) = E^{\mu\nu}_{\quad\beta} \tmnd \quad , \quad \p^{\beta}V_{\beta\rho} =0 \quad , \quad V=
V_{\mu}^{\,\,\mu} = 0 \quad . \label{VT}\ee

\no The above properties of the nonsymmetric tensor
$V_{\beta\rho}(T)$ are due to the transverse nature of the
operator $E^{\mu\nu\rho}$ and the property (\ref{t2})
respectively. We can rewrite (\ref{let}) as follows

\be {\cal L}\left[e,T\right] = - \left[ e_{\beta\rho} +
V_{\beta\rho}(T)\right]\left[ e^{\rho\beta} +
V^{\rho\beta}(T)\right] - c\, e^2 + {\cal L}_{C}\left[T\right]
\quad . \label{let2} \ee

\no Since $V_{\beta\rho}(T)$ is traceless, after the shift
$e_{\beta\rho} \to \tilde{e}_{\beta\rho} - V_{\beta\rho}$ we have
two  non-dynamic terms for $\tilde{e}_{\beta\rho}$ decoupled from
${\cal L}_{C}\left[T\right]$. The fields $\tilde{e}_{\beta\rho}$
can thus be trivially integrated out in the path integral. We
conclude that the particle content of ${\cal L}\left[e,T\right]$
is the same one of Curtright's theory (\ref{lc}), i.e., one
massive spin-2 particle. Notice that $e_{\beta\rho}$ does not need
to be traceless to be shifted.

On the other hand,  instead of integrating over $e_{\beta\rho}$ we
can Gaussian  integrate over $\tmnd$. We end up with a dual
massive Lagrangian ${\cal L}^*_m(e) = {\cal L}\left[e,T(e)\right]$
where

\bea \tmnd (e) &=& T_{\mu}(e) \eta_{\rho\nu}-
T_{\nu}(e)\eta_{\rho\mu} - \frac{1}{6\, m^2} \left\lbrack
2E_{\mu\nu\beta}e_{\rho}^{\,\,\beta} +
E_{\rho\nu\beta}e_{\mu}^{\,\,\beta} +
E_{\mu\rho\beta}e_{\nu}^{\,\,\beta} \right\rbrack \label{eom2}\\
T_{\mu}(e) &=& \frac 1{4 m^2}E_{\mu\alpha\beta}e^{\alpha\beta}
\label{eom3} \eea

 The tensor $\tmnd (e)$ is obtained from the equations of motions of (\ref{let}) and satisfies (\ref{t2}).
The Lagrangian density ${\cal L}_m^*(e)$  must describe a massive
spin-2 particle with 5 propagating degrees of freedom. Other
interesting features of ${\cal L}_m^*(e)$ can be anticipated from
(\ref{let}). Since the last three terms of (\ref{let}) can only
generate, after Gaussian integration, kinetic terms of second
order in derivatives of $e_{\mu\nu}$ it is already clear that we
can have a massive spin-2 particle without necessarily a
Fierz-Pauli mass term which corresponds to $c=-1$. Given that
$\tmnd$ is coupled to $e^{\rho\beta}$ via
$V_{\beta\rho}(T)e^{\rho\beta}$, due to the properties (\ref{VT})
the kinetic terms (mass independent terms) will be invariant under
the linearized reparametrizations and Weyl transformations:

\be \delta e_{\alpha\beta} = \p_{\beta}\xi_{\alpha} +
\eta_{\alpha\beta} \, \phi
 \label{gt1} \ee

\no In the special case $c=-1/4$ the whole massive theory is
invariant under the Weyl transformations $\delta_W e_{\alpha\beta}
= \eta_{\alpha\beta} \phi $. Explicitly, after a redefinition
$e_{\alpha\beta} \to m\, e_{\alpha\beta}/\sqrt{2}$, the Gaussian
integrals over the mixed symmetry tensors furnish a massive dual
model for arbitrary values of $c$ which is our main result, i.e.,

\bea {\cal L}^*_m &=& -\frac 12
\p^{\mu}e^{(\alpha\beta)}\p_{\mu}e_{(\alpha\beta)} +
\left[\p^{\alpha}e_{(\alpha\beta)} \right]^2- \frac 13
\left(\p^{\alpha}e_{\alpha\beta}\right)^2 - \frac{m^2}2 \left(
e_{\alpha\beta}e^{\beta\alpha} + c\, e^2 \right) \nn \\
&+& \frac 16 \p^{\mu}e\p_{\mu}e - \frac 13
\p^{\alpha}e_{\alpha\beta}\p^{\beta}e \label{ldualm} \eea

\no where $e_{(\alpha\beta)}=\left( e_{\alpha\beta} +
e_{\beta\alpha}\right)/2$ and $e_{\left[\alpha\beta\right]}=\left(
e_{\alpha\beta} - e_{\beta\alpha}\right)/2$. The reader can check
that the mass independent terms of (\ref{ldualm}) are indeed
invariant under (\ref{gt1}). At this point one might try to bring
the arbitrary mass term in (\ref{ldualm}) to the Fierz-Pauli form
with $c=-1$ by means of a local change of variables $e_{\mu\nu} =
\tilde{e}_{\mu\nu} + \frac a2 \,\tilde{e}\eta_{\mu\nu}$ by tuning
the real constant $a$ conveniently without affecting the kinetic
terms which are Weyl invariant. However, this is not always
possible. Explicitly we have

\be e_{\mu\nu}e^{\nu\mu} + c\, e^2 \to
\tilde{e}_{\mu\nu}\tilde{e}^{\nu\mu} + \tilde{c}\, \tilde{e}^2
\label{aat} \ee

\no where $\tilde{c} = c + (1+4c)(a + a^2)$. We have three classes of mass terms according to  $c < -1/4 $, $c >
-1/4$ and the fixed point $c=-1/4$. Except for the fixed point, any representative of a class is continuously
connected to anyone else of the same class by varying the parameter $a$. No interclass jump is allowed. Thus,
without loss of generality we can simply pick up for instance $c=-1$, $c=0$ and $c=-1/4$, where the first case
corresponds to the usual Fierz-Pauli mass term while the other two cases can not be brought into the FP form.

Defining the massless action $S_{m=0}^*=\int d^4x \, {\cal
L}^*_{m=0}$, the equations of motion of the massive dual model
(\ref{ldualm}) can be written as

\be m^2\, \left( e_{\beta\alpha} + c \, \eta_{\beta\alpha}\, e
\right) = K_{\alpha\beta} \label{eq1} \ee

\no with the massless Killing tensor given by

\bea K_{\alpha\beta} = \frac{\delta S_{m=0}^*}{\delta
e^{\alpha\beta}} &=& \frac{\Box}{2}\left( e_{\alpha\beta} +
e_{\beta\alpha}\right) - \frac 12
\p_{\delta}\left(\p_{\alpha}e_{\beta}^{\,\,\,\delta} +
\p_{\beta}e_{\alpha}^{\,\,\,\delta}\right) +
\frac{\eta_{\alpha\beta}}3\left(\p_{\mu}\p_{\nu}e^{\mu\nu} - \Box
e\right) \nn \\
 &-&\frac 12 \p_{\beta}\p_{\delta}e^{\delta}_{\,\,\alpha} + \frac
16 \p_{\alpha}\p_{\mu}e^{\mu}_{\,\,\beta} + \frac 13
\p_{\alpha}\p_{\beta} e \label{k} \eea

\no Due to the symmetries  (\ref{gt1}) we have
$\eta^{\alpha\beta}K_{\alpha\beta}=0=\p^{\beta}K_{\alpha\beta} $.
Thus, from $\eta^{\alpha\beta}$ and $\p^{\beta}$ on (\ref{eq1}) we
have

\bea e &=& 0 \,\, , \label{tss1} \\
\p^{\beta}e_{\alpha\beta} &=& 0 \,\, , \label{trans1}
 \eea

\no Although (\ref{tss1}) only holds for $c\ne -1/4$, it can be
implemented as a gauge condition of the Weyl symmetry if $c
=-1/4$. So we assume (\ref{tss1}) and (\ref{trans1}) henceforth
for all values of $c$. Moreover, the antisymmetric part of
(\ref{eq1}) now leads to

\be e_{\alpha\beta} - e_{\beta\alpha} = 0 \,\, , \label{anti1} \ee

\no Therefore, although $e_{\left[\alpha\beta\right]}$ appear
under derivatives in ${\cal L}_m^*$, they play the role of
auxiliary fields. Finally, (\ref{eq1}) becomes the Klein-Gordon
equations:

\be \left(\Box - m^2 \right) e_{\alpha\beta} = 0 \,\, .
\label{kg1} \ee

The equations (\ref{tss1})-(\ref{kg1}) are the Fierz-Pauli
conditions. They guarantee that we have 5 propagating degrees of
freedom corresponding to a massive spin-2 particle for any value
of the constant $c$.

We have also checked unitarity by calculating the two point
amplitude $A_2(k)$. Introducing arbitrary sources $T_{\mu\nu}$ we
have

\be e^{\int d^4 k \, A_2(k)} = \int {\cal D}e_{\alpha\beta}e^{i\,
S_m^*\left[e\right] + i \int d^4 x \,
e_{\alpha\beta}T^{\alpha\beta} } \label{zj} \ee

\no From (\ref{zj}) it can be shown that $A_2(k)$ is given in
terms of the saturated propagator in momentum space as follows

\be A_2 (k) = -i
\left(T^{\mu\nu}(k)\right)^*\left[G^{-1}_{\mu\nu\alpha\beta}(k)\right]T^{\alpha\beta}(k)
\quad . \label{a2k} \ee

\no where $T_{\mu\nu}^*(k)$ is the complex conjugated of the
Fourier transform of the sources. In general, in the massive
theory there are no constraints on the source except at $c=-1/4$
where the source must be traceless due to the Weyl symmetry. The
propagator in momentum space can be obtained ($G^{-1}(k)=G^{-1}(\p
\to i\, k)$) from the differential operator below in coordinate
space, we have suppressed the four indices for simplicity,

\bea G^{-1} &=& \frac{P_{SS}^{(2)}}{\Box - m^2}
-\frac{2}{m^4}\left(\frac{\Box}{3} + m^2 \right) P_{SS}^{(1)} -
\frac{2}{m^4}\left(\frac{\Box}{3} - m^2 \right) P_{AA}^{(1)} \nn\\
&-& \frac{ \Box}{3\, m^4} \left[ P_{AS}^{(1)} + P_{SA}^{(1)}\right] + \frac{P_{AA}^{(0)}}{m^2} - \frac{1+c}{
m^2\left(4c+1\right)}P_{SS}^{(0)} - \frac{1+ 3c}{m^2 \left(4c+1\right)}P_{WW}^{(0)} \nn\\ &+&
\frac{\sqrt{3}}{m^2\left( 4c +1\right)}\left[ P_{WS}^{(0)} + P_{SW}^{(0)}\right]\label{gmenos1} \eea

\no The spin-s projection operators $P_{JJ}^{(s)}$ and the transition operators $P_{IJ}^{(s)} \, , \, I \ne J $
are given in the appendix. They satisfy the simple algebra

\be P_{IJ}^{(s)}P_{KL}^{(r)} = \delta^{sr}\delta_{JK} P_{IL}^{(s)}
\quad . \label{algebra} \ee

\no We have used (\ref{algebra}) in order to obtain
(\ref{gmenos1}) by inverting\footnote{There is no inverse at
$c=-1/4$ due to the Weyl symmetry however, in this case we can add
a gauge fixing term $-\lambda \, e^2$ which amounts to substitute
$c\to -1/4 + \lambda $ in (\ref{gmenos1}).}
$G_{\alpha\beta\mu\nu}$ which on its turn is defined by $S_m^*
=\int d^4 x {\cal L}_m^* = \int d^4x
h^{\alpha\beta}G_{\alpha\beta\mu\nu}h^{\mu\nu} $. The only pole in
(\ref{gmenos1}) occurs in the spin-2 sector at $\Box = m^2$  just
like in the usual Fierz-Pauli theory where $c=-1$. The calculation
of  the imaginary part of the residue ($R_m$) of $A_2(k)$ at
$k^2=-m^2$ proceeds in the same way as in the massive FP theory.
The fact that $T_{\mu\nu}$ is not symmetric in our case does not
make any difference since $P_{SS}^{(2)}$ projects out in the
symmetric, transverse and traceless sector anyway. Namely,

\bea R_m &=& \lim_{k^2 \to -m^2} (k^2 + m^2) \Im m \left[
A_2(k)\right] =
\left(T^{\mu\nu}\right)^*\left[P_{SS}^{(2)}\right]_{\mu\nu\alpha\beta}T^{\alpha\beta}\nn\\
&=& \left(T_{TT}^{\mu\nu}\right)^*\left(T_{TT}\right)_{\mu\nu} =
\sum_{i,j} \vert T^{ij}_{TT} \vert^2 > 0 \label{rm} \eea

\no where the symmetric, transverse and traceless tensor is given
by
$T_{TT}^{\mu\nu}=\left(P_{SS}^{(2)}\right)^{\mu\nu\alpha\beta}T_{\alpha\beta}
$ and we have used $T_{TT}^{0\mu} =0 $ which follows from
$k_{\alpha}T_{TT}^{\alpha\beta} =0 $ in the frame
$k_{\alpha}=\left(m,0,0,0\right)$. Thus, our massive dual model
${\cal L}_m^*$ is free of ghosts and describes one spin-2 massive
particle for arbitrary real values of $c$.

For a closer comparison with the usual massive FP model it is
instructive to decompose $e_{\mu\nu}$ into symmetric and
antisymmetric parts. From (\ref{ldualm}) we have

\bea {\cal L}^*_m &=& -\frac 12
\p^{\mu}h^{(\alpha\beta)}\p_{\mu}h_{(\alpha\beta)} + \frac 18
\p^{\mu} h \p_{\mu}h - \frac 12 \left( \p^{\alpha}B_{\alpha\mu}
\right)^2 + \frac{m^2}2 B_{\alpha\beta}^2 \nn \\
&+& \frac 23 \left( \p^{\alpha}h_{\alpha\mu} - \frac 12 \p^{\alpha}B_{\alpha\mu} - \frac 14 \p_{\mu}h \right)^2
- \frac{m^2}2 \left(h_{\alpha\beta}^2 + c\, h^2 \right)\label{ldualm2} \eea

\no where

\be e_{\mu\nu} = h_{\mu\nu} + B_{\mu\nu}  \quad , \quad
h_{\mu\nu}=h_{\nu\mu} \quad , \quad B_{\mu\nu} = - B_{\nu\mu}
\quad . \label{ahb} \ee

\no For convenience we write down the usual Fierz-Pauli theory:

\be {\cal L}^{FP}_m = -\frac 12
\p^{\mu}h^{(\alpha\beta)}\p_{\mu}h_{(\alpha\beta)} + \frac 14
\p^{\mu} h \p_{\mu}h + \left(\p^{\alpha}h_{\alpha\beta} - \frac 12
\p_{\beta}h \right)^2 - \frac{m^2}2 \left(h_{\alpha\beta}^2 - \,
h^2 \right) \label{lfpm} \ee

\no The vectors inside the large parenthesis in Eq.
(\ref{ldualm2}) and Eq. (\ref{lfpm}) are related to harmonic
gauges (de Donder gauge) for the massless theory to be discussed
in the next section, see (\ref{gc2}).

We see in (\ref{ldualm2}) that the coupling between $B_{\mu\nu}$
and $h_{\mu\nu}$ is nontrivial and can not be undone by means of
any local field redefinition. In fact the functional integral over
$B_{\mu\nu}$ leads to a nonlocal effective action for the
symmetric field $h_{\mu\nu}$. Moreover if we simply set
$B_{\mu\nu}=0$ the remaining theory is no longer ghost free.

The absence of ghosts in (\ref{ldualm2}) is surprisingly not only
because of the coupling between $B_{\mu\nu}$ and $h_{\mu\nu}$, see
\cite{pvn73}, but also because of the non Fierz-Pauli mass term
(for $c \ge -1/4$). Another example of ghost-free
symmetric-antisymmetric coupling has been found recently  in
\cite{ms12} where the mass term must be of the usual Fierz-Pauli
type ($c=-1$). Their kinetic terms do not contain the trace $h$
and can not be brought to the form appearing in (\ref{ldualm}) or
in (\ref{lfpm}) by local transformations.

At this point we comment on another work in the literature. In
\cite{zino} one also finds a first-order master action depending
on a mixed symmetry tensor $\omega_{\mu\left[ \nu\alpha\right] }$
and $e_{\mu\nu}=h_{\mu\nu} + B_{\mu\nu}$, with an arbitrary real
constant $a$ in the mass term similar to (\ref{let}).
Symbolically, the Lagrangian of \cite{zino} can be written as

\be {\cal L}_I = \omega \cdot \omega + \omega \left(\p h + \p
B\right) - \frac{m^2}2\left( h_{\mu\nu}^2 - h^2\right) + a
B_{\mu\nu}^2 \quad . \label{zino} \ee

\no Integrating over $\omega_{\mu\nu\alpha}$ one gets the usual
massive Fierz-Pauli action \cite{fp} displayed in (\ref{lfpm})
plus $a B_{\mu\nu}^2 $. Therefore, the antisymmetric field
$B_{\mu\nu}$ does not play any physical role and vanishes on shell
which is similar to the trace $e$, see (\ref{tss1}), in our model
(\ref{ldualm}). Thus, (\ref{zino}) describes one massive spin-2
particle for any value of $a$, similar to (\ref{ldualm}).  The
case $a=0$ is special since we have a gauge symmetry $\delta
B_{\mu\nu} = \Lambda_{\left[\mu\nu\right]} \, , \, \delta
\omega_{\mu\left[\nu\alpha\right]} = \p_{\mu}
\Lambda_{\left[\nu\alpha\right]}$, see \cite{ahs}. This is the
analogue of the $c=-1/4$ case in our model (\ref{ldualm}). In
general, after the change of variables $\omega_{\mu\left[
\nu\alpha\right] } \to \omega_{\mu\left[ \nu\alpha\right] } +
\p_{\mu}B_{\nu\alpha} $ the field $B_{\mu\nu}$ disappears from
(\ref{zino}) except from the last term. This shows that there is
no physical coupling between $B_{\mu\nu}$ and $\omega_{\mu\left[
\nu\alpha\right] }$ just like in (\ref{let}) where the trace $e$
does not couple to $T_{\left[\lambda\sigma\right]\beta} $. So the
arbitrariness in both master actions (\ref{ldualm}) and
(\ref{zino}) is related to degrees of freedom which are physically
decoupled from the dual field. An important difference between
(\ref{ldualm}) and the theory obtained from (\ref{zino}) after
integration over $\omega_{\mu\nu\alpha}$ is the surprisingly
absence of ghosts in (\ref{ldualm}) as compared to the latter
case.

 Finally, if we compare the usual massless Fierz-Pauli theory (linearized Einstein-Hilbert)
 to the $m\to 0$ limit of our massive dual model by calculating $A_2(k)$
saturated with symmetric conserved sources, then the same mass
discontinuity problems found in \cite{vdv,zak} for the usual
massive Fierz-Pauli theory show up. From (\ref{gmenos1}) we see
that only the spin-2 sector can lead to long range interactions.
It gives rise to the same result for tree level interacting
potential as the massless limit of the massive Fierz-Pauli theory.
The remaining (lower spin) terms are contact terms which may be
neglected.

\section{The massless case}

According to the dualization procedure summarized in the introduction of the last section it is expected that
the $m=0$ case be singular somehow. At $m=0$ the functional integral over $e_{\mu\nu}$ leads to a constraint on
the spin connection $\omega_{\mu a b}$ instead of quadratic kinetic terms. So the particle content of (\ref{lc})
does not need to reproduce the massless FP theory which describes a massless spin-2 particle. Indeed, it can be
proved, see \cite{bfrt}, that ${\cal L}_C(m=0)$ contains no particle at all\footnote{By using the first-order
dual formulation (\ref{let}) at $m=0$ an alternative proof which is explicitly covariant and gauge independent
can be done \cite{ds2}.}. Since we have used a similar dualization procedure in deriving ${\cal L}_m^*$, the
particle content of ${\cal L}_{m=0}^*$ is not known ${\it a \, priori}$.

First, we note that ${\cal L}_{m=0}^*$ is invariant under gauge
transformation which act also in the antisymmetric part of
$e_{\mu\nu}$ thus enlarging (\ref{gt1}), namely,

 \be \delta e_{\alpha\beta} = \p_{\beta}\xi_{\alpha} +
\eta_{\alpha\beta} \, \phi +
\epsilon_{\alpha\beta\mu\nu}\p^{\mu}\Lambda^{\nu}
 \label{gt2} \ee

\no In terms of dual fields $B_{\mu\nu}^* =
\epsilon_{\mu\nu\alpha\beta}B^{\alpha\beta}$ we can write $\delta
B_{\mu\nu}^* = \p_{\left[ \mu\right.}\Lambda_{\left.\nu\right]} $.
The importance of this type of symmetry in order to avoid ghosts
in nonsymmetric tensor theories has been emphasized in
\cite{ddmc93}.

We fix the gauge in the antisymmetric sector imposing

\be \epsilon^{\mu\nu\alpha\beta}\p_{\nu}e_{\alpha\beta} =0 \quad .
\label{gc1} \ee

\no Notice that the antisymmetric gauge transformation is
reducible under $\delta \Lambda_{\mu} = \p_{\mu}\Phi $.
Consequently, we can only fix three independent degrees of freedom
in agreement with the transverse gauge condition (\ref{gc1}).

Regarding the reparametrization and Weyl symmetry, we have found
convenient to choose harmonic gauges (like the Lorentz gauge in
electrodynamics and the de Donder gauge $\p^{\mu}h_{\mu\nu} -
\p_{\mu}h/2=0$ for symmetric tensors) which have residual gauge
invariances under harmonic functions $\Box \xi = 0 = \Box \phi$.
Respectively, we define the gauges

\bea G_{\beta} = \p^{\alpha}e_{\alpha\beta} + 3\,
\p^{\alpha}e_{\beta\alpha} -
\p_{\beta}e &=& 0 \quad , \label{gc2} \\
H = \p^{\alpha}\p^{\beta}e_{\alpha\beta} - \Box e &=& 0 \quad .
\label{gc3} \eea

\no Each of the gauge conditions (\ref{gc1}),(\ref{gc2}) and
(\ref{gc3})  breaks only one of the three symmetries (\ref{gt2}).
Now we can define the gauge fixed action (disregarding the
decoupled Faddeev-Popov term) and obtain an invertible operator
$G_{\mu\nu\alpha\beta}$ as follows

\be S = \int d^4 x \left[ {\cal L}_{m=0}^*(e) + \lambda_1
G_{\mu}G^{\mu} + \lambda_2 \left(
\epsilon^{\mu\nu\alpha\beta}\p_{\nu}e_{\alpha\beta} \right)^2 +
\lambda_3 H^2 \right] = \int d^4 x
e_{\mu\nu}G^{\mu\nu\alpha\beta}e_{\alpha\beta} \label{gf} \ee

\no After expanding $G_{\mu\nu\alpha\beta}$ on the basis of spin-s
operators $P_{IJ}^{(s)}$ given in the appendix we obtain the
inverse operator (suppressing indices)

\bea G^{-1} &=& \frac{2 P_{SS}^{(2)}}{\Box} -\frac{1}{18\lambda_1\Box}\left\lbrack (1-12\lambda_1)P_{SS}^{(1)} +
(1-48\lambda_1)P_{AA}^{(1)} + (1+24\lambda_1) \left( P_{AS}^{(1)} + P_{SA}^{(1)}\right)\right\rbrack \nn\\ &-&
\frac{P_{AA}^{(0)}}{2\lambda_2\Box} + \frac{P_{SS}^{(0)}}{3\lambda_3\Box^2}  + \frac{\lambda_1 - \lambda_3
\Box}{9\lambda_1\lambda_3\Box^2} P_{WW}^{(0)} - \frac{\sqrt{3}\left[ P_{WS}^{(0)}+
P_{SW}^{(0)}\right]}{9\lambda_3\Box^2} \label{gmenos2} \eea

\no We have a gauge independent massless pole in the spin-2 and
spin-1 sectors. Next we deduce the constraints on the sources due
to the gauge symmetries (\ref{gt2}) and calculate the residue in
$A_2(k)$ at $k^2=0$. From the invariance of the source term under
(\ref{gt2}):

\be \int d^4 x T^{\mu\nu} \, \delta e_{\mu\nu} = 0 \quad .
\label{dte} \ee

\no We deduce

\bea T_{\mu\nu} &=& T_{(\mu\nu)} + \p_{\mu}T_{\nu}-\p_{\nu}T_{\mu}
\label{sc1} \\ \eta_{\mu\nu}T^{(\mu\nu)} &=& 0 \quad , \quad
\p_{\nu}T^{\mu\nu} = 0 \quad . \label{sc2} \eea

\no In particular, we have $\omega_{\mu\nu}T^{\mu\nu} =0 =
\theta_{\mu\nu}T^{\mu\nu}$. Defining the shorthand notation

\be
T_{\mu\nu}^*\left(P_{IJ}^{(s)}\right)^{\mu\nu\alpha\beta}T_{\alpha\beta}
\equiv T^*P_{IJ}^{(s)}T \quad , \label{shorth} \ee

\no it is not difficult to check that the spin-0 operators drop
out from the saturated propagator:

\be T^* P_{IJ}^{(0)} T =0 \quad . \label{spin0} \ee

\no In the spin-1 and spin-2 sectors we have

\be T^* P_{SS}^{(1)} T = \frac 12 T^* \omega \, T = T^*
P_{AA}^{(1)}
 T \quad , \quad T^* \left[ P_{AS}^{(1)} + P_{SA}^{(1)} \right] T
 = - T^* \omega \, T \quad , \label{tomegat1} \ee
\be T^* P_{SS}^{(2)} T = T^*_{(\mu\nu)}T^{(\mu\nu)} - \frac 12 T^*
\omega\, T \quad . \label{tomegat2} \ee

 \no where

 \be T^* \omega \, T = T_{\mu\nu}^* \frac{k^{\mu}k_{\alpha}}{k^2}
 T^{\alpha\nu} \quad . \label{tomegat} \ee

 \no Collecting all the above results in the formula (\ref{a2k}) for $A_2(k)$, the gauge
 parameters $\lambda_j \, , \, j=1,2,3$ cancel out and we have the
 gauge independent result

 \be A_2(k) = \frac{2\, i}{k^2}\left\lbrack
 T^*_{(\mu\nu)}T^{(\mu\nu)} + T^* \omega \, T \right\rbrack \quad .
 \label{a2k0} \ee

\no From the momentum space expressions $T_{\mu\nu} = T_{(\mu\nu)} +
i\left(k_{\mu}T_{\nu}-k_{\nu}T_{\mu}\right)$ and $k_{\nu}T^{\mu\nu} =0 $ we can deduce $k_{\mu}T^{\mu\nu} = 2\,
i\left(k^2 T^{\nu} - k^{\nu} k_{\nu}T^{\nu} \right)$. Consequently,

\be T^* \omega \, T = 4 \left\lbrack k^2 T_{\mu}^* T^{\mu} - \vert
k_{\nu}T^{\nu} \vert^2\right\rbrack \quad . \label{tomegatk} \ee

\no Since we are interested in the residue at $k^2=0$ we stick
from now on to the light-like frame $k_{\mu}=(k,k,0,0)$. From
$k_{\nu}T^{\mu\nu} = 0 $ we have $T^{(\mu\nu)}k_{\nu} =
i\left\lbrack k^2 T^{\mu} -\, k^{\mu}\left( k_{\alpha}T^{\alpha}
\right)\right\rbrack $ which in the above frame  leads to the four
equations

\bea T^{(02)} &=& -T^{(21)} \quad , \quad T^{(03)} = -T^{(31)} \quad , \label{j02} \\
T^{00} &=& T^{11} + 2\, i \, k_{\alpha}T^{\alpha} \quad , \quad
T^{(01)} = - T^{11} - \, i \, k_{\alpha}T^{\alpha} \quad .
\label{j00} \eea

\no It follows from (\ref{j02}) and (\ref{j00}) that

\be T_{(\mu\nu)}^*T^{(\mu\nu)} = \vert T^{22}\vert^2 + \vert
T^{33}\vert^2 + 2\vert T^{23}\vert^2 + 2 \vert
k_{\alpha}T^{\alpha}\vert^2 \quad . \label{tt} \ee

\no From (\ref{tomegatk}) at $k^2=0$ and (\ref{tt}) we finally
obtain for the imaginary part of the residue of $A_2(k)$ at
$k^2=0$:

\be R_0 = 2\left( \vert T_{22} \vert^2 + \vert T_{33} \vert^2 +
2\vert T_{23} \vert^2 -  2 \vert k_{\alpha}T^{\alpha}\vert^2
\right) = \vert T_{22} - T_{33} \vert^2 + 4 \vert T_{23} \vert^2
> 0 \label{ro} \ee

\no where we have used $ k_{\alpha}T^{\alpha} = \left( T_{22} +
T_{33}\right)/(2\, i) $ which follows from the first equation in
(\ref{j00}) and the traceless condition $T_{00}- T_{11} = T_{22} +
T_{33}$.

In summary, $R_0 > 0$ and  the massless theory is ghost free.

Next we check the equations of motion coming from ${\cal L}_{m=0}^* $ at the gauge conditions
(\ref{gc1}),(\ref{gc2}) and (\ref{gc3}). Those equations correspond to $K_{\mu\nu}=0$, see (\ref{k}). First, the
antisymmetric part $K_{\left[\mu\nu\right]}=0$ leads to

\be \p_{\mu}\left(\p^{\alpha}e_{\alpha\nu} \right) -
\p_{\nu}\left(\p^{\alpha}e_{\alpha\mu} \right) = 0 \quad
\Rightarrow \p^{\alpha}e_{\alpha\nu} = \p_{\nu} \Phi  \quad .
\label{phi} \ee

\no where $\Phi$ is so far an arbitrary scalar field. Back in
$\p^{\mu}G_{\mu}=0$ , $H=0$ and $G_{\mu}=0$ we have

\be \Box e = 0 \quad , \quad \p^{\mu}\p^{\nu}e_{\mu\nu} = \Box
\Phi = 0 \label{boxe} \ee

\be 3 \p^{\nu}e_{\mu\nu} = \p_{\mu}\left(e-\Phi\right) \label{22}
\ee

\no Back in $K_{\mu\nu}=0$ we deduce $ \Box e_{(\mu\nu)} = 0 $.
Now we can define the field

\be h_{\mu\nu} = e_{(\mu\nu)} - \frac{\eta_{\mu\nu}}3\left(\frac
e2 + \Phi \right) \quad . \label{hmn} \ee

\no which satisfies

\bea h_{\left[\mu\nu\right]}&=&0 = \p^{\mu}h_{\mu\nu} \quad , \quad \Box h_{\mu\nu} = 0  \quad , \label{h1}\\
h &=& \frac 13 \left(e- 4\Phi \right) \quad . \label{h2} \eea

\no All the equations written so far are invariant under residual reparametrization and Weyl gauge
transformations with harmonic parameters ($\Box \xi_{\mu} =0 = \Box \phi $). Since they imply

\bea \delta h_{\mu\nu} &=& \frac 12 \left\lbrack \p_{\mu}\xi_{\nu} + \p_{\mu}\xi_{\nu} - \eta_{\mu\nu} \p \cdot
\xi  \right\rbrack  \label{rh} \\ \delta \Phi &=& \phi + \p \cdot \xi \quad , \quad \delta e = 4 \phi +  \p
\cdot \xi \quad , \label{ra} \eea

\no we can use the residual Weyl invariance to get rid of the scalar field $\Phi $ imposing:

\be \Phi - e = 0 \label , \label{phia} \ee

\no Since (\ref{phia}) is reparametrization invariant, no further
requirement is made on the harmonic reparametrization  parameters
$\xi_{\mu}$ which can thus, be used to get rid of extra four
degrees of freedom of $h_{\mu\nu}$. So $h_{\mu\nu}$ contains, see
(\ref{h1}), two helicity states $\pm 2$.

\no Regarding the antisymmetric part $e_{\left[\mu\nu\right]}$,
since the solution of the gauge condition (\ref{gc1}) leads to a
field strength of some vector field:

\be e_{\mu\nu}- e_{\nu\mu} = \p_{\mu} A_{\nu} - \p_{\nu} A_{\mu}
\quad . \label{nl}\ee

\no If we plug it back in
$\p^{\mu}\left(e_{\mu\nu}-e_{\nu\mu}\right) = - \p_{\nu} h$ we
have

\be \Box A_{\mu} - \p_{\mu}\left( \p \cdot A  \right) = -\p_{\mu} h \quad . \label{max} \ee

\no We can split the general solution of (\ref{max}): $A_{\mu} = A
_{\mu}^{max} + A _{\mu}^{h}$. Where $A _{\mu}^{max}$ is a general
solution of Maxwell equations $\p^{\mu}F_{\mu\nu}(A^{max}) = 0$
while $A _{\mu}^{h}$ is a specific solution of the non-homogeneous
equation (\ref{max}). Clearly, $A _{\mu}^{h}$ does not represent
an independent degree of freedom. Moreover we can get rid of $A
_{\mu}^{max}$ by using a constrained symmetry of the massless
model. Namely, from (\ref{ldualm2}) we see that the massless
theory depends on $B_{\mu\nu}$ only through the combination
$\p^{\mu}B_{\mu\nu}$ which is invariant under $\delta
B_{\mu\nu}=\epsilon_{\mu\nu\alpha\beta}\p^{\alpha}\Lambda^{\beta}
+ \p_{\mu}C_{\nu}- \p_{\nu}C_{\mu} $ where the gauge parameter
$\Lambda^{\mu}$ is arbitrary while $C_{\mu}$ must satisfy the free
Maxwell equations $\p^{\mu}F_{\mu\nu}(C) = 0$. Since the gauge
condition (\ref{gc1}) does not impose any further constraint on
the parameters $C_{\mu}$ they can be used to cancel $A
_{\mu}^{max}$. Thus, the antisymmetric part
$e_{\mu\nu}-e_{\nu\mu}$ does not contribute to the spectrum of the
theory which consists only of one massless spin-2 particle.

There is another way of checking the particle content of ${\cal
L}_{m=0}^*$. In fact\footnote{We thank an anonymous referee for
calling our attention to \cite{cmu2}}, ${\cal L}^*_{m=0}$ has
appeared before in \cite{cmu2} in a completely different way via
solution of a constraint in a massless master action. As pointed
out in a footnote in \cite{bch}, it is useful to rewrite ${\cal
L}_{m=0}^*$ with help of a non-dynamical vector field $v_{\mu}$ as
follows,

\bea {\cal L}^*_{m=0}&=&  {\cal L}_0(h) + \frac 13 v^2 + \frac 23 v^{\mu}\left(\p^{\alpha}B_{\alpha\mu}
+ \p^{\alpha}h_{\alpha\mu}\right) \nn \\
&+& h_{\mu\nu}T^{(\mu\nu)} + 2\, B_{\mu\nu} \p^{\mu} T^{\nu} \quad . \label{lv0} \eea

\no where we have added sources satisfying the constraints
(\ref{sc1}) and (\ref{sc2}) and

\be {\cal L}_0(h) = -\frac 12 \p^{\mu}h^{(\alpha\beta)}\p_{\mu}h_{(\alpha\beta)} + \frac 16 \p^{\mu} h \p_{\mu}h
+ \left( \p^{\mu}h_{\mu\nu}\right)^2 - \frac 13 \p^{\mu}h_{\mu\nu}\p^{\nu}h \label{l0} \ee

\no If we integrate over $B_{\mu\nu}$ in the generating functional
we get a functional delta function which enforces a constraint
whose general solution is $v_{\mu} = 3 T_{\mu} + \p_{\mu} \psi $
where $\psi $  is an arbitrary scalar field. Then, after
integrating over $v_{\mu}$ we get an effective theory containing
$h_{\mu\nu}$ and $\psi$. It turns out that after the redefinition
$h_{\mu\nu} \to \tih_{\mu\nu} - \eta_{\mu\nu} \left( \psi + \tih
\right) $ the scalar field $\psi $ disappears and we end up with
the linearized Einstein-Hilbert theory with a modified source
term, i.e.,

\be {\cal L}_{eff}^* = {\cal L}^{FP}_{m=0}(\tih) +
\tih_{\mu\nu}\tilde{T}^{\mu\nu} \label{lv1} \ee

\no where ${\cal L}^{FP}_{m=0}$ corresponds to (\ref{lfpm}) at
$m=0$ and

\be \tilde{T}_{\mu\nu} = T_{(\mu\nu)} - \p_{\mu}T_{\nu} -
\p_{\nu}T_{\mu} + 2\, \eta_{\mu\nu} \p \cdot T \quad
.\label{nsource} \ee

\no The new source is symmetric and conserved
$\p^{\mu}\tilde{T}_{\mu\nu} = 0$ thanks to the constraint
$\p_{\mu}T^{(\mu\nu)} =\p_{\mu}F^{\mu\nu}(T)$ which follows from
(\ref{sc1}) and (\ref{sc2}).

In conclusion, as in
 the usual massless FP theory (linearized Einstein-Hilbert)., the symmetric field $h_{\mu\nu}$ couples to a
symmetric and conserved source, see more comments in \cite{cmu2}.
Although we have found a pole in both spin-1 and spin-2 sectors of
the propagator, there is only one spin-2 massless particle in the
spectrum. The residue calculation is somehow similar to the
linearized Einstein-Hilbert (LEH) theory whose propagator contains
a gauge independent massless pole in the spin-2 and also in the
spin-0 sectors but there is only one spin-2 particle in the
spectrum. Indeed, if we saturate the LEH propagator with the
effective source (\ref{nsource}) and calculate the residue $R_0=
\tilde{T}^* \left\lbrack 2 P_{SS}^{(2)} - P_{SS}^{(0)}
\right\rbrack \tilde{T} $ we have exactly the same result of
formula (\ref{ro}).

Since we have been able to get rid of the antisymmetric field
$B_{\mu\nu}$ via local field redefinitions, one might try the same
manipulations in the massive case. In fact, we can still trade the
antisymmetric field $B_{\mu\nu}$ in a vector field $v_{\mu}$,
i.e., if we add $ m^2 B_{\mu\nu}^2 $ to the right-hand side of
(\ref{lv0}), after integrating over $B_{\mu\nu}$ we have a
Maxwell-Proca theory for the vector field $v_{\mu}$. However, if
we further integrate over $v_{\mu}$ we end up with a nonlocal
action for the symmetric field $h_{\mu\nu}$ as we have mentioned
in section 2. Thus, in the massive case the antisymmetric field is
not simply an auxiliary field.

\section{Conclusion}

Here we have shown that unitarity does not lead to a unique
description of massive spin-2 particles in terms of a rank-2
tensor in $D=4$. In particular, the mass term does not need to fit
in the widely used\footnote{One exception is \cite{dpr} where the
mass of the ``would be'' ghost is not set to infinity ($c=c(\Box)
\ne -1$). The theory is ghost free due to a phenomenological
reason. Namely, the mass of the ghost lies presumably above the
energy scale below which the massive gravitational theory is
supposed to work. Another exception is \cite{zino}, see comment at
the end of section 2.} Fierz-Pauli form $e_{\mu\nu}e^{\nu\mu} + c
\, e^2$ with $c=-1$ or more generally in the Fierz-Pauli class $c
< -1/4$. The other two classes $c>-1/4$ and $c=-1/4$ also lead to
ghost-free theories. The arbitrariness in the mass term is related
to the absence of a totally antisymmetric part
($T_{\left[\left[\mu\nu\right]\rho\right]}=0$) for the mixed
symmetry tensor of the dual theory.

An important ingredient in our model is the use of a nonsymetric
tensor $e_{\mu\nu}$ which naturally appears in the flat space
limit of first-order formulations of gravity. Another important
point is the nontrivial coupling between the symmetric and
antisymmetric parts of $e_{\mu\nu}$ which is also present in the
recent suggestion of \cite{ms12} where the mass term must be of
the usual Fierz-Pauli form.

Unfortunately, the mass discontinuity \cite{vdv,zak} at $m \to 0$
is independent of the arbitrariness in the mass term and coincides
(up to contact terms) with the Fierz-Pauli result. However, since
we have some freedom in the mass term which does not need to be
fine tuned as in the Fierz-Pauli theory, one may hope of solving
the discontinuity problem by adding non-linear terms without
necessarily creating ghosts.

Regarding the massless theory (section 3), it has appeared before
in \cite{cmu2} via a different procedure. Its spectrum consists of
a massless spin-2 particle. Thanks to the three gauge symmetries
(\ref{gt2}) the model is ghost-free. The antisymmetric part
$e_{\left\lbrack \mu \nu \right\rbrack}$ can be eliminated and we
end up with the usual linearized Einstein-Hilbert theory with
modified sources.

\section{Appendix}

From the spin-1 and spin-0 projection operators acting on vector
fields, respectively,

\be \theta_{\mu\nu} = \eta_{\mu\nu} -
\frac{\p_{\mu}\p_{\nu}}{\Box} \quad , \quad \omega_{\mu\nu} =
\frac{\p_{\mu}\p_{\nu}}{\Box} \quad , \label{pvectors} \ee

\no one \cite{pvn73} can build up projection and transition
operators mentioned in the text. First we present the symmetric
operators

\be \left( P_{SS}^{(2)} \right)^{\lambda\mu}_{\s\s\alpha\beta} =
\frac 12 \left( \theta_{\s\alpha}^{\lambda}\theta^{\mu}_{\s\beta}
+ \theta_{\s\alpha}^{\mu}\theta^{\lambda}_{\s\beta} \right) -
\frac{\theta^{\lambda\mu} \theta_{\alpha\beta}}{D-1} \quad ,
\label{ps2} \ee

\be \left( P_{SS}^{(1)} \right)^{\lambda\mu}_{\s\s\alpha\beta} =
\frac 12 \left(
\theta_{\s\alpha}^{\lambda}\,\omega^{\mu}_{\s\beta} +
\theta_{\s\alpha}^{\mu}\,\omega^{\lambda}_{\s\beta} +
\theta_{\s\beta}^{\lambda}\,\omega^{\mu}_{\s\alpha} +
\theta_{\s\beta}^{\mu}\,\omega^{\lambda}_{\s\alpha}
 \right) \quad , \label{ps1} \ee

\be \left( P_{SS}^{(0)} \right)^{\lambda\mu}_{\s\s\alpha\beta} =
\frac 1{D-1} \, \theta^{\lambda\mu}\theta_{\alpha\beta} \quad ,
\quad \left( P_{WW}^{(0)} \right)^{\lambda\mu}_{\s\s\alpha\beta} =
\omega^{\lambda\mu}\omega_{\alpha\beta} \quad , \label{psspww} \ee

\be \left( P_{SW}^{(0)} \right)^{\lambda\mu}_{\s\s\alpha\beta} =
\frac 1{\sqrt{D-1}}\, \theta^{\lambda\mu}\omega_{\alpha\beta}
\quad , \quad  \left( P_{WS}^{(0)}
\right)^{\lambda\mu}_{\s\s\alpha\beta} = \frac 1{\sqrt{D-1}}\,
\omega^{\lambda\mu}\theta_{\alpha\beta} \quad , \label{pswpws} \ee

\no They satisfy the symmetric closure relation

\be \left\lbrack P_{SS}^{(2)} + P_{SS}^{(1)} +  P_{SS}^{(0)} +
P_{WW}^{(0)} \right\rbrack_{\mu\nu\alpha\beta} =
\frac{\eta_{\mu\alpha}\eta_{\nu\beta} +
\eta_{\mu\beta}\eta_{\nu\alpha}}2 \quad . \label{sym} \ee

\no The remaining antisymmetric and mixed symmetric-antisymmetric
operators are given by

\be \left( P_{AA}^{(1)} \right)^{\lambda\mu}_{\s\s\alpha\beta} =
\frac 12 \left(
\theta_{\s\alpha}^{\lambda}\,\omega^{\mu}_{\s\beta} -
\theta_{\s\alpha}^{\mu}\,\omega^{\lambda}_{\s\beta} -
\theta_{\s\beta}^{\lambda}\,\omega^{\mu}_{\s\alpha} +
\theta_{\s\beta}^{\mu}\,\omega^{\lambda}_{\s\alpha}
 \right) \quad , \label{paa1} \ee

\be \left( P_{SA}^{(1)} \right)^{\lambda\mu}_{\s\s\alpha\beta} =
\frac 12 \left(
\theta_{\s\alpha}^{\lambda}\,\omega^{\mu}_{\s\beta} +
\theta_{\s\alpha}^{\mu}\,\omega^{\lambda}_{\s\beta} -
\theta_{\s\beta}^{\lambda}\,\omega^{\mu}_{\s\alpha} -
\theta_{\s\beta}^{\mu}\,\omega^{\lambda}_{\s\alpha}
 \right) \quad , \label{pas1} \ee

\be \left( P_{AS}^{(1)} \right)^{\lambda\mu}_{\s\s\alpha\beta} =
\frac 12 \left(
\theta_{\s\alpha}^{\lambda}\,\omega^{\mu}_{\s\beta} -
\theta_{\s\alpha}^{\mu}\,\omega^{\lambda}_{\s\beta} +
\theta_{\s\beta}^{\lambda}\,\omega^{\mu}_{\s\alpha} -
\theta_{\s\beta}^{\mu}\,\omega^{\lambda}_{\s\alpha}
 \right) \quad , \label{psa1} \ee

\be \left( P_{AA}^{(0)} \right)^{\lambda\mu}_{\s\s\alpha\beta} =
\frac 12 \left( \theta_{\s\alpha}^{\lambda}\theta^{\mu}_{\s\beta}
- \theta_{\s\alpha}^{\mu}\theta^{\lambda}_{\s\beta} \right) \quad
, \label{paa0} \ee

\no They satisfy the antisymmetric closure relation (see appendix
B of \cite{arias})

\be \left\lbrack P_{AA}^{(1)} + P_{AA}^{(0)}
\right\rbrack_{\mu\nu\alpha\beta} =
\frac{\eta_{\mu\alpha}\eta_{\nu\beta} -
\eta_{\mu\beta}\eta_{\nu\alpha}}2 \quad . \label{asym} \ee

\no Adding up (\ref{sym}) and (\ref{asym}) we have

\be \left\lbrack P_{SS}^{(2)} + P_{SS}^{(1)} +  P_{SS}^{(0)} +
P_{WW}^{(0)} +  P_{AA}^{(1)} + P_{AA}^{(0)}
\right\rbrack_{\mu\nu\alpha\beta} =
\eta_{\mu\alpha}\eta_{\nu\beta} \quad . \label{full} \ee

\no The reader can check that the operators satisfy the simple
algebra

\be P_{IJ}^{(s)}P_{KL}^{(r)} = \delta^{sr}\delta_{JK} P_{IL}^{(s)}
\quad . \label{algebra2} \ee

\section{Acknowledgements}

 We thank Antonio S. Castro, Alvaro de S.
Dutra, J\'ulio M.Hoff da Silva, Marcelo B. Hott and Elias L.
Mendon\c ca  for discussions. This work is partially supported by
CNPq.

\end{document}